\documentclass[11pt]{article}

\usepackage{graphicx}
\usepackage{ifpdf}
\usepackage[a4paper,bindingoffset=0.2in,left=0.5in,right=0.5in,top=1in,bottom=1in,footskip=.25in]{geometry}

\ifpdf
   \DeclareGraphicsExtensions{.pdf,.png,.jpg,.mps}
   \usepackage{hyperref}
\else
\fi

\usepackage[space]{grffile}
\usepackage{latexsym}
\usepackage{amsfonts,amsmath,amssymb}
\usepackage{url}
\usepackage{fancyref}
\usepackage{hyperref}
\hypersetup{colorlinks=false,pdfborder={0 0 0},}
\usepackage{color}
\usepackage{caption}
\usepackage{authblk}
\usepackage{setspace}
\usepackage{lineno}
\usepackage{gensymb}
\newcommand{\beginsupplement}{%
        \setcounter{table}{0}
        \renewcommand{\thetable}{S\arabic{table}}%
        \setcounter{figure}{0}
        \renewcommand{\thefigure}{S\arabic{figure}}%
     }

\long\def\/*#1*/{} 

\begin{document}

\title{The network architecture of value learning}
\author[1]{Marcelo G. Mattar}
\author[1]{Sharon L. Thompson-Schill}
\author[2,3]{Danielle S. Bassett\thanks{Corresponding author: dsb@seas.upenn.edu}}
\affil[1]{Department of Psychology, University of Pennsylvania, Philadelphia, PA 19104, USA}
\affil[2]{Department of Bioengineering, University of Pennsylvania, Philadelphia, PA 19104, USA}
\affil[3]{Department of Electrical \& Systems Engineering, University of Pennsylvania, Philadelphia, PA 19104, USA}

\date{ }

\renewcommand\Authands{ and }

\maketitle

\clearpage
\section*{Abstract}

Value guides behavior. With knowledge of stimulus values and action consequences, behaviors that maximize expected reward can be selected. Prior work has identified several brain structures critical for representing both stimuli and their values. Yet, it remains unclear how these structures interact with one another and with other regions of the brain to support the dynamic acquisition of value-related knowledge. Here, we use a network neuroscience approach to examine how BOLD functional networks change as 20 healthy human subjects learn the values of novel visual stimuli over the course of four consecutive days. We show that connections between regions of the visual, frontal, and cingulate cortices become increasingly stronger as learning progresses, and that these changes are primarily confined to the temporal core of the network. These results demonstrate that functional networks dynamically track behavioral improvement in value judgments, and that interactions between network communities form predictive biomarkers of learning.

\clearpage

\section*{Introduction}

A human's behavior is fundamentally driven by their existing notions of value \cite{simon1955behavioral}. From a vast repertoire of possible actions, humans choose ones that have either been reinforced through prior rewards, or have the potential to bring reward in the future \cite{shizgal1997neural,montague2006imaging}. The concept of value is foundational to decision making, allowing for diverse alternatives to be placed on a common scale, thereby facilitating choices that maximize expected reward. While a notion of value is intrinsic to many stimuli (e.g. a red apple appears more valuable than a brown apple), in many cases it must be learned through experience (e.g. coffee is more valuable than expected given its appearance). Conceptual representations of value can be acquired through trial and error (right vs. wrong), as well as through learning of declarative information \cite{squire1992declarative,packard2002learning,delgado2012reward}.

The representation of the value of objects requires the engagement of systems that represent object information and systems that represent value. Visual objects, and their form in particular, are represented throughout the occipital and temporal lobes, occupying part of what is known as the \emph{visual system} \cite{felleman1991distributed,grill2004human}. Stimulus values, on the other hand, are represented primarily in subcortical and medial frontal areas, in a collection of structures referred to as the \emph{valuation system} \cite{bartra2013valuation}. Notably, identifying a stimulus and retrieving its value requires the concerted engagement of both systems. 

Numerous studies have attempted to elucidate the specific functions of each region in these systems using clever task designs and sophisticated methodological approaches \cite{grill2004human, bartra2013valuation, de2008fine, grill2014functional, cohen2005functional}. Indeed, the success of these studies is evident by the sheer number of compartmentalized structures identified along with their associated functions. In the valuation system, for example, basal ganglia structures respond in proportion to reward prediction errors \cite{packard2002learning,o2004dissociable,abler2006prediction} -- a crucial signal in feedback-based learning, while frontal regions of the valuation system respond in proportion to the actual values of stimuli \cite{o2004reward,bartra2013valuation} -- a signal important for value-based decision making. However, a fundamental gap in our knowledge lies in delineating how the different cortical and subcortical areas composing the valuation and visual systems interact with one another and with other regions of the brain to allow effective behavioral choices built on the computations of and comparisons between stimulus values. Knowledge of these patterns of interaction is an essential step in turning from a compartmentalized or modular view of brain function, towards a more integrative and dynamic notion of how different regions cooperate to subserve behaviors as complex as perception and decision making. 

Here we address this gap by taking an explicit \emph{network neuroscience} perspective. This novel analytical framework has encountered great success in characterizing how learning modulates the patterns of statistical dependencies between regional activities, bridging and relating descriptions at the microscale with emerging architectures at the mesoscale \cite{bassett2013task,bassett2015learning}. In a cohort of 20 healthy adult human subjects, we examine how the pattern of functional interactions between brain regions change during the learning of monetary values of novel visual stimuli. With over 1500 trials completed across four consecutive days of practice, participants learned the monetary values of 12 rendered three-dimensional shapes through feedback consisting either of the value of the shape selected (10 subjects) or the correctness of the shape selected (10 subjects). 

We hypothesized that learning accompanies gradual changes in the architecture of functional networks, with progressively increased integration between visual and valuation systems to allow proficient task performance. We further hypothesized that the recruitment of network regions and modules varies depending on the type of feedback received in the task, with an increased engagement of basal ganglia structures for participants learning by trial and error and an increased engagement of visual structures for participants learning declarative information. We hypothesized that this effect is strongest in the early stages when learning is more pronounced and a full representation of the stimulus values is incomplete. Finally, following recent empirical \cite{bassett2013task,bassett2015learning} and theoretical \cite{fedorenko2014reworking} evidence, we posited that learning primarily modulates functional connectivity between regions in the temporal core of a dynamic network. 

Our results not only confirm these hypotheses, but they also suggest that functional networks can almost perfectly distinguish whether a novel subject has already learned the stimulus values, and can significantly classify the type of feedback received in the learning protocol. These findings demonstrate that behavioral improvements in value judgment are represented in patterns of functional connectivity that change in a characteristic manner predictive of learning. Moreover, our study offers a set of novel analytical approaches applicable to the study of human learning specifically and dynamic networks more broadly.

\clearpage
\section*{Results}

\subsection*{Experimental paradigm}

\begin{figure}[!htbp]
\centering
\includegraphics[]{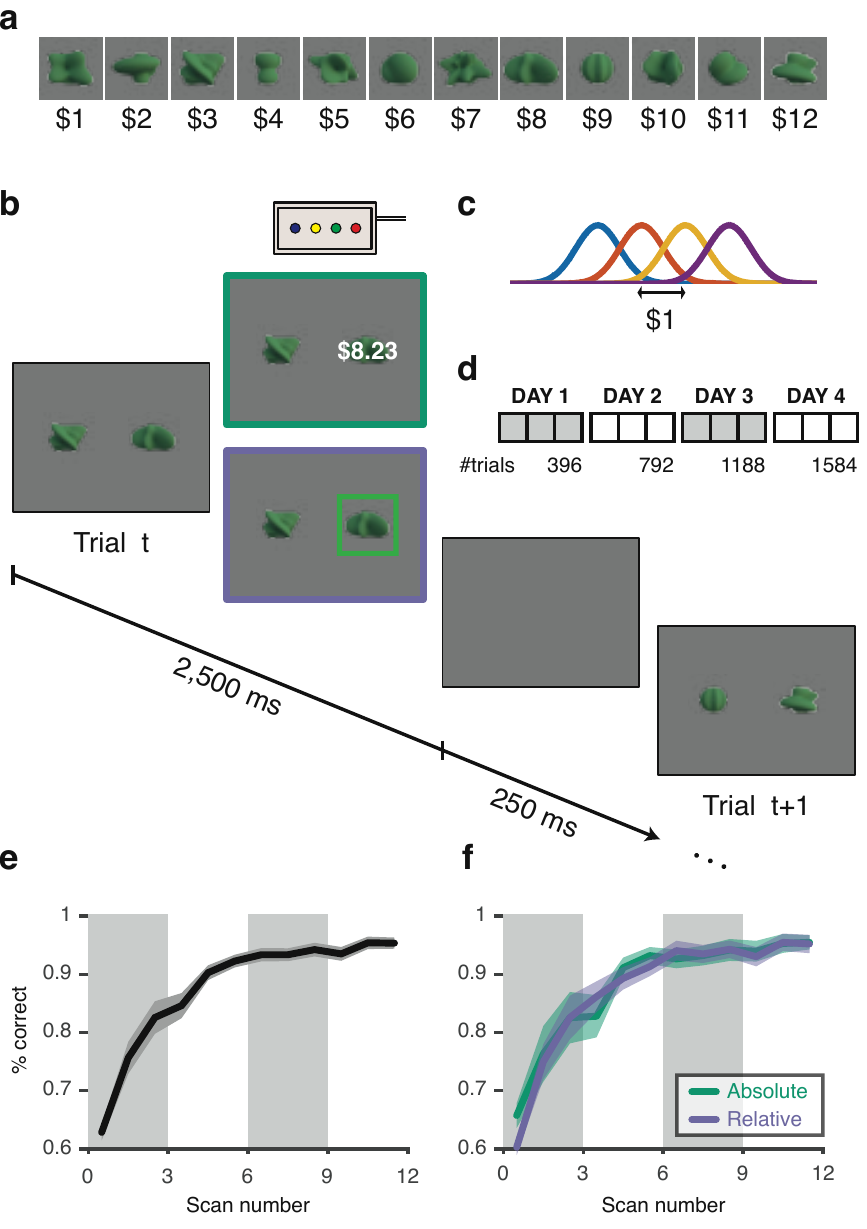}
\caption{\textbf{Experimental paradigm and behavioral results.}
\emph{(a)} Stimulus set and corresponding values. Twelve abstract shapes were computer generated, and an integer value between \$1 and \$12 was assigned to each. This value remained constant over the four days of training.
\emph{(b)} Task paradigm. Participants were presented with two shapes side by side on the screen and asked to choose the shape with the higher monetary value. Once a selection was made, either the value of the shape selected (\emph{absolute feedback}) or the correctness of the selection (\emph{relative feedback}) was provided as feedback. Each trial lasted 2.75 s (250 ms inter-stimulus interval).
\emph{(c)} On each trial, the empirical value of each shape was drawn from a Gaussian distribution with fixed mean (i.e., the true value), as described in panel \emph{(a)}, and standard deviation of \$0.50.
\emph{(d)} The experiment was conducted over four consecutive days, with three experimental scans (396 trials) on each day, for a total of 1584 trials.
\emph{(e)} Participants’ accuracy in selecting the shape with higher \emph{expected} value improved steadily over the course of the experiment, increasing from chance level in the first few trials to approximately 95\% in the final few trials.
\emph{(f)} Task accuracy followed a similar profile for all participants in both the \emph{absolute} feedback and \emph{relative} feedback groups.
\label{fig1}}
\end{figure}

Twenty healthy adult human subjects learned the monetary value of 12 novel visual stimuli over the course of four consecutive days (Fig.~\ref{fig1}a). On each trial of the experiment, participants selected which of two shapes simultaneously present on the screen had the highest value, after which they received feedback based on their response. Participants were randomly assigned to two groups determining the type of feedback that they would receive. Ten participants received \emph{relative} feedback: a green (red) square surrounding the selected shape, signaling correct (incorrect) response. The other ten participants received \emph{absolute} feedback: the value of the selected shape was presented to the participant, but not the value of the non-selected shape (Fig.~\ref{fig1}b). Although each shape had a true value, the empirical value used for each trial was drawn from a Gaussian distribution with a fixed mean (i.e., true value; Fig.~\ref{fig1}a,c) and with a standard deviation of \$0.50. We arrived at this latter choice by performing preliminary behavioral studies to identify a standard deviation value that led to quantitatively similar learning rates in both the relative and absolute feedback groups.

We collected blood oxygen level dependent (BOLD) functional MRI data from each participant as they performed the task. A total of 12 scan runs over 4 days were completed by each person (three scans per session), totaling 1584 trials (Fig.~\ref{fig1}d). The average accuracy in selecting the shape with the highest mean value at each trial gradually improved over the course of the experiment, increasing from approximately 50\% (chance) in the first few trials to approximately 95\% in the final few trials. This behavioral improvement was consistent across participants and between feedback groups (Fig.~\ref{fig1}e,f). Two participants from each group were excluded on the basis of head motion and task performance (Fig. S1). 

\subsection*{Evolution of functional networks throughout learning}

\begin{figure}[!htbp]
\centering
\includegraphics[]{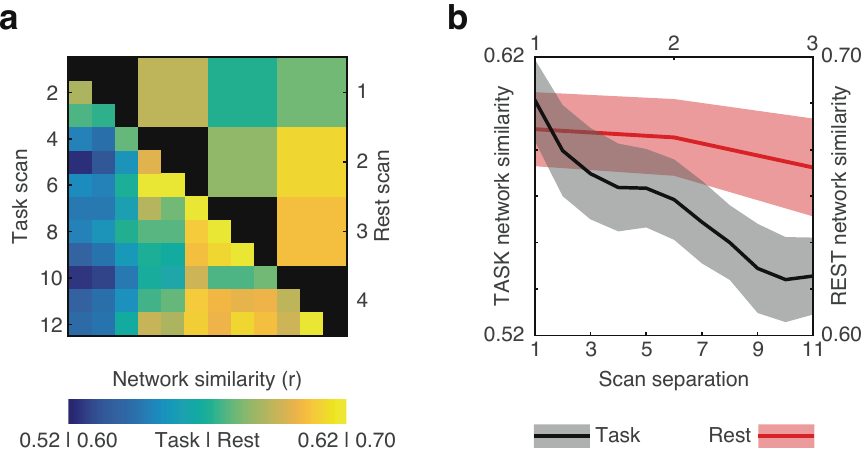}
\caption{\textbf{Functional networks underlying task execution evolve slowly over the course of learning.}
\emph{(a)} Lower diagonal: Network similarity (Pearson correlation coefficient) between functional connectivity matrices corresponding to each pair of task scans. Upper diagonal: Network similarity (Pearson correlation coefficient) between functional connectivity matrices corresponding to each pair of rest scans, conducted over the same period of time as the task scans. 
\emph{(b)} Average network similarity as a function of temporal separation between corresponding scans. Black line: task scans; Red line: rest scans.
\label{fig2}}
\end{figure}

Using this value learning task, we first tested whether global changes in functional connectivity occurred concurrently with changes in behavior \cite{medaglia2015cognitive}. We created functional networks (or graphs) for each participant by subdividing their gray matter volume into $N=112$ cortical and subcortical areas (nodes) and calculating the statistical dependency between the BOLD activity time courses from each pair of nodes (edges) \cite{bullmore2011brain}. We defined one such functional network for each of the 12 scans, and we represented that network as an $N \times N$ weighted adjacency matrix. Then, for each pair of scans, we calculated the Pearson correlation coefficient between their associated pair of matrices, intuitively measuring the inter-scan similarity in the pattern of statistical dependencies between regional BOLD time series. We observed that functional networks were more similar to each other when the corresponding scans were close to one another in time than when they were far from one another in time (Fig.~\ref{fig2}a). Interestingly, functional networks tended to evolve most from the first to the second day (scans 1--3 and 4--6), which was also the period that saw the greatest improvement in accuracy (Fig.~\ref{fig1}e). 

We summarized these results by calculating the average functional network similarity (Pearson correlation between adjacency matrices) as a function of scan separation (Fig.~\ref{fig2}b). A repeated-measures analysis of variance revealed that average network similarity significantly decreased as scan separation increased, $F(10, 150) = 11.43, P<0.001$. As a critical null model, we considered the baseline resting state scans acquired prior to each task session, in order to rule out potential effects at the session level that were unspecific to learning. A repeated-measures analysis of variance did not yield significant results, suggesting that average functional network similarity at rest did not significantly decrease as rest scan separation increased, $F(2, 30) = 0.70, P=0.51$.  Similar results were obtained when separately considering the two feedback groups (\emph{absolute}: $F(10,70) = 3.46, P=0.0010$ for task vs. $F(2,14) = 0.33, P=0.73$ for rest; \emph{relative}: $F(10,70) = 10.51, P<0.001$ for task vs. $F(2,14) = 0.48, P=0.63$ for rest, Fig. S2). Moreover, similar results were also obtained when considering the average functional networks for each day as opposed to each scan, yielding an equal number of time points for the task and rest data ($F(2,30) = 7.51, P=0.0023$, Fig. S2). These results indicate that functional networks evolve steadily during task execution, suggesting their sensitivity to the learning of value information. Importantly, a homogeneous change in functional connectivity in the entire network cannot account for these results (e.g. an overall increase or decrease in connectivity), given that a correlation measure discounts a mean offset. Therefore, these results demonstrate that the pattern of edge weights changes throughout learning, with distinct connections changing in different ways. We thus next turned our attention changes at a finer scale to effectively characterize the reconfiguration in network architecture associated with learning.

\subsection*{Relationship between behavioral and network changes}

\begin{figure}[!htbp]
\centering
\includegraphics[]{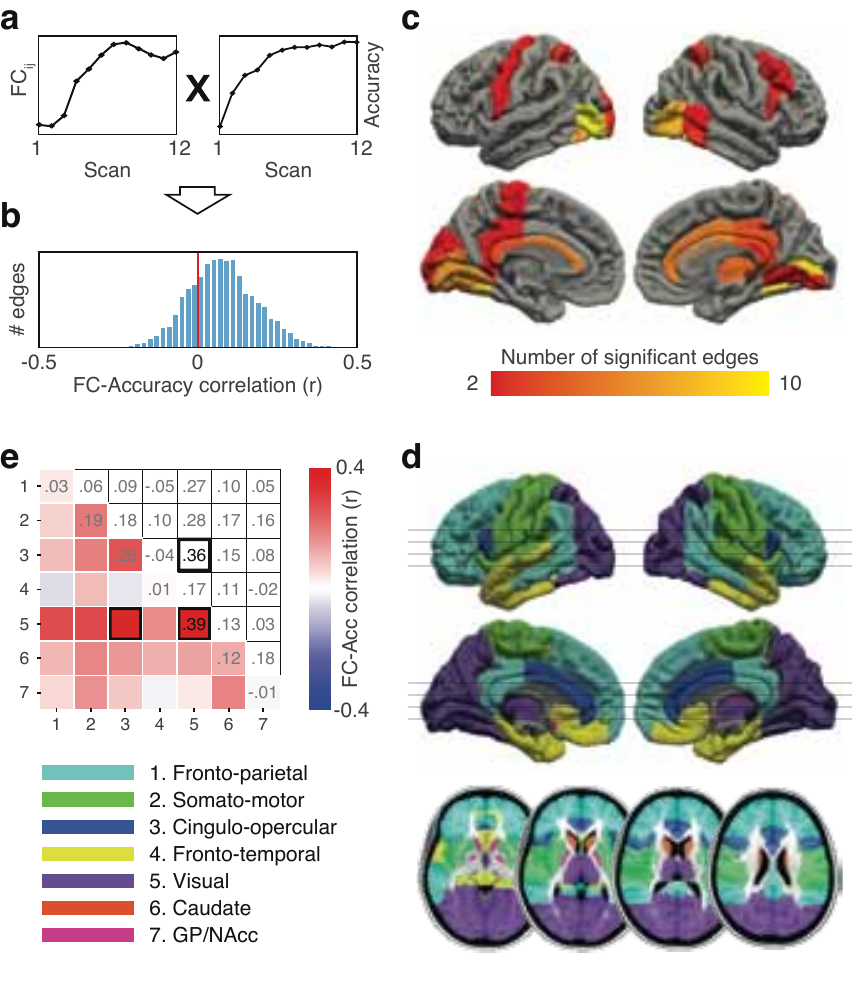}
\caption{\textbf{Changes in functional connectivity track changes in task accuracy.}
\emph{(a)} Approach for identifying network dynamics associated with behavior. Left inset: The strength of functional connectivity for an example edge connecting nodes $i$ and $j$. Right inset: Task accuracy for the same participant.
\emph{(b)} Histogram of Pearson correlation coefficients between edge weight and task accuracy. Edge weights are, on average, positively correlated with task accuracy, indicating that edge weights tend to increase as learning progresses.
\emph{(c)} Number of edges departing from each node whose variation over time correlated significantly with changes in behavior (Bonferroni corrected at $\alpha=0.05$). Connections associated with learning primarily involved regions in the visual cortex, but to a lesser degree also regions in cingulate, somato-motor, and dorso-lateral frontal areas.
\emph{(d)} Community structure. Colors represent sets of regions that displayed coherent activity with one another during task execution, across all participants and task sessions. Color code is displayed in the legend in the bottom-left of the figure. Subcortical structures are displayed in representative axial slices, shown at the bottom.
\emph{(e)} Correlation between average edge weight within or between communities and task accuracy. Communities whose interactions varied in a correlated manner with task accuracy (Bonferroni corrected at $\alpha=0.05$) are highlighted in the matrix. Community order is displayed in the legend in the bottom-left of the figure.
\label{fig3}}
\end{figure}

Having established that functional networks evolve steadily over the course of learning, we next turned to an examination of functional changes at the scales of nodes and edges. To quantify the degree to which variation in edge weight over time was associated with learning, we calculated the Pearson correlation coefficient between each participant's task accuracy and the strength of functional connectivity at each edge in the network, across scans (Fig.~\ref{fig3}a). We observed a tendency for correlation values to be positive ($M = 0.079, SD = 0.11$), suggesting that functional connectivity on average increases as task accuracy increases (Fig.~\ref{fig3}b). Similarly, we confirmed that, across subjects, the global network strength was significantly correlated with behavior ($M = 0.21 \pm 0.077$ (SEM), one-sample $t$-test on Fisher normalized correlation values: $t(15) = 2.73, P=0.015$). These results suggest that behavioral improvements in this task are associated with a global increased coherence in BOLD activity.

Despite an overall tendency for functional connectivity values to increase over the course of learning, this profile was not present in all edges of the network (Fig.~\ref{fig3}b). Thus, we wished to distinguish the parts of the network that were related to behavioral changes from those that did not change or from those that changed in a manner unrelated to behavior. Using a nonparametric permutation-based approach, we compared the mean correlation coefficient between task accuracy and edge weight, across participants, with a null distribution generated by randomly permuting the order of the scans. We then applied Bonferroni correction for multiple comparisons ($112 \times 111/2 = 6216$ tests) and counted the number of edges departing from each node whose changes across time were significantly correlated with improvements in task accuracy. We observed that regions of the visual cortex -- in particular around the calcarine sulcus, inferior lateral occipital cortex, and posterior fusiform -- included the most edges whose weights tracked learning. Following these strongest hubs of behaviorally-linked connections were additional regions in cingulate, somato-motor, and dorso-lateral frontal cortices (Fig.~\ref{fig3}c).

While these results provide information about focal regions from which important edges emanate, they do not address the question of which edges specifically change strength in concert with task accuracy. To examine this finer-scale structure while maintaining interpretability, we categorized edges grouped by the corresponding cognitive systems recruited by value learning. We used a data-driven approach based on a network-based clustering method to uncover these cognitive systems or \emph{functional modules}. Specifically, we identified groups of brain regions that displayed coherent activity with one another during task execution, forming network communities. Using this approach, we obtained subject-specific communities for every trial block, and obtained a representative consensus partition by identifying the groups of nodes that were consistently assigned to the same community across participants and across time. The consensus partition divides the brain into seven distinct communities: a fronto-parietal community spanning regions of the dorso-lateral, ventro-lateral, and ventro-medial frontal cortices, posterior cingulate, and inferior parietal lobe; a somato-motor community comprised of regions in the precentral and postcentral gyri and sulci; a cingulo-opercular community covering the anterior cingulate and frontal operculum; a visual community composed of the occipital, posterior parietal, and inferior temporal cortices and thalamus; and two subcortical communities: one formed by bilateral caudate, and one formed by the nucleus accumbens and globus pallidus (Fig.~\ref{fig3}d). 

We then used this community structure -- largely in agreement with known divisions of the visual and value networks -- to summarize groups of network connections that robustly changed with learning. First, we calculated the average edge weight within each community or between each pair of communities, and then we calculated the correlation between these module-level estimates of functional interactions and task accuracy (Fig.~\ref{fig3}e). To assess the significance of these effects, we compared the average (Fisher-Z transformed) correlation value, across subjects, with a null distribution of 10,000 correlation values obtained by permuting the order of the scans uniformly at random. Two community-level interactions showed significant correlations with task accuracy (Bonferroni corrected at $\alpha=0.05$): as task accuracy increased, connection strength similarly increased (i) within the visual community (adjusted $P$-value: $p < 0.0028$); and (ii) between visual and cingulo-opercular communities (adjusted $P$-value: $P=0.011$). Similar results were obtained when analyzing aboslute and relative feedback groups separately (Fig. S3). Together, these results suggest that interactions between visual and cingulo-opercular networks change in accordance with task accuracy.

\subsection*{Predicting a person's learning stage from their functional connectivity pattern}

\begin{figure}[!htbp]
\centering
\includegraphics[]{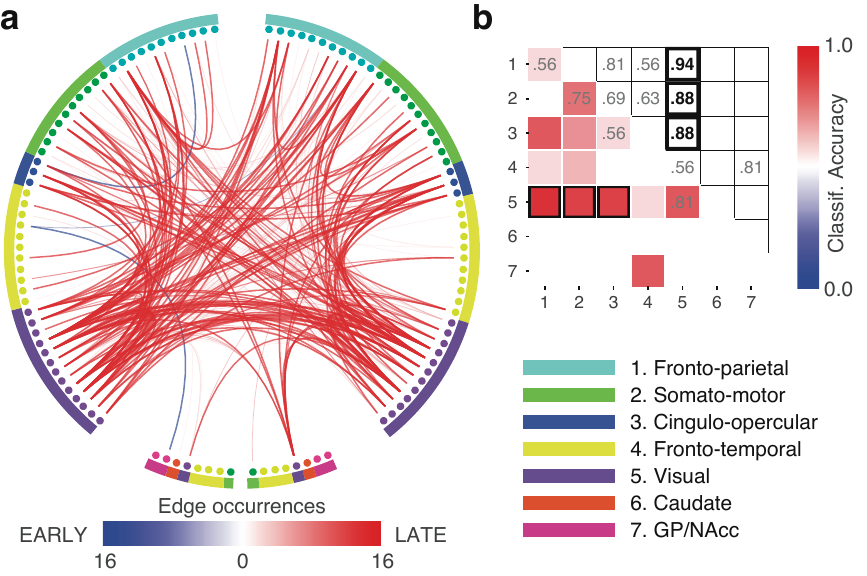}
\caption{\textbf{Functional connectivity between network modules predicts a person's learning stage.}
\emph{(a)} Using a cross-validation procedure, the subset of edges where a one-sample $t$-test yielded a test statistic with $P$-value lower than $0.001$ was selected as a \emph{predictive network}. The figure shows the number of cross-validation folds in which each edge was identified as part of the predictive network. Edges whose strength correlated positively with task accuracy are displayed in red. Edges whose strength correlated negatively with task accuracy are displayed in blue. 
\emph{(b)} The same procedure repeated separately for each pair of communities. Cell colors and numbers represent classification accuracy in labeling held-out data as coming from scans early or late in the learning process. Communities whose interactions classified held-out data significantly (Bonferroni corrected at $\alpha=0.05$) are highlighted. Community order is displayed in the bottom right.
\label{fig4}}
\end{figure}

We next turned to the stricter test of out-of-sample prediction. Specifically, we set out to test whether snapshots of an unseen participant's network could be correctly classified as coming from early \emph{versus} late stages in the learning process. We used a \emph{leave-one-out} cross-validation procedure to select predictive network edges from each set of $n-1$ subjects, and we compared their strengths in the left-out, independent participant's data. For each cross-validation fold, predictive network edges were those in which a one-sample $t$-test across subjects demonstrated a significant correlation with behavior at $P<0.001$. The number of edges present in the predictive networks ranged from 180 to 341 ($M = 222.6, SD = 36.6$). We then calculated the average strength in this network for the first three scans (day 1) and for the last three scans (day 4) of the left-out participant, and we classified the dataset with lower (higher) strength as early (late). Using this approach, we were able to correctly classify 15 out of the 16 participant's data (accuracy: $93.75\%$ vs. chance: $50\%$; one-tailed binomial test: $P<0.001$). The edges that appeared most frequently in the predictive network linked the visual community with itself (purple) and with the fronto-parietal (cyan), cingulo-opercular (blue) and somato-motor (green) communities (Fig.~\ref{fig4}a). 

For a baseline comparison, using a predictive network composed of all (6216) possible edges we were able to correctly classify data from 12 out of the 16 participants (accuracy: $75\%$ vs. chance: $50\%$; one-tailed binomial test: $P=0.038$). The performance of this classifier is a consequence of the overall increase in the average edge strength of the network and not due to the exact pattern of edge strengths. We controlled for this global effect by removing the average network strength at each scan and repeating the classification procedure, forcing the predictive networks to incorporate only edges whose variation with respect to other edges (and not with respect to a baseline) predicted learning session. Using this alternative cross-validation procedure, the number of edges present in the predictive networks ranged from 105 to 154 ($M = 132.5, SD = 13.7$) and classification was correct in all 16 participants (accuracy: $100\%$ vs. chance: $50\%$; one-tailed binomial test: $P<0.001$; Fig. S4). These results confirm that the unique pattern of edge strengths is a robust predictor of a participant's learning stage.

To gain insight into the specific modules that enable this classification, we repeated these analyses separately for each pair of communities, selecting each time the edges connecting the two communities for which a one-sample $t$-test across subjects demonstrated a significant correlation with behavior at $p<0.001$ (Fig.~\ref{fig4}b). Using a Bonferroni correction for multiple comparisons ($\alpha=0.05$), the data from the left-out participant was significantly classified above chance (50\%) when the predictive network was comprised of edges connecting: (i) visual and fronto-parietal modules (accuracy: $93.75\%$; one-tailed binomial test, adjusted $P$-value: $P=0.0054$); (ii) visual and somato-motor modules (accuracy: $87.50\%$; one-tailed binomial test, adjusted $P$-value:  $P=0.044$); and (iii)  visual and cingulo-opercular modules (accuracy: $87.50\%$; one-tailed binomial test, adjusted $P$-value:  $P=0.044$). Together, these results confirm that interactions between visual and fronto-cingulate regions not only track behavioral improvements, but can also be used to significantly determine the amount of training that an unseen person has completed.

\subsection*{Insights into fundamental constraints on dynamic network architecture}

We have thus far described the specific network components at various levels that change in concert with learning of value. But how do these components relate to the mesoscale architecture of the network? To gain insight into this question, we asked whether regions and edges that track task performance were present in the temporal \emph{core} of the network, thought to be critical for task-specific processes \cite{bassett2013task}, or in the temporal \emph{periphery} of the network, thought to be critical for domain-general processes \cite{fedorenko2014role}. We propose that learning primarily modulates functional connectivity between stiff regions in the temporal core of a dynamic network, in light of converging empirical evidence \cite{bassett2013task,bassett2015learning} and given the high specificity of the processes involved in this task.

\begin{figure}[!htbp]
\centering
\includegraphics[]{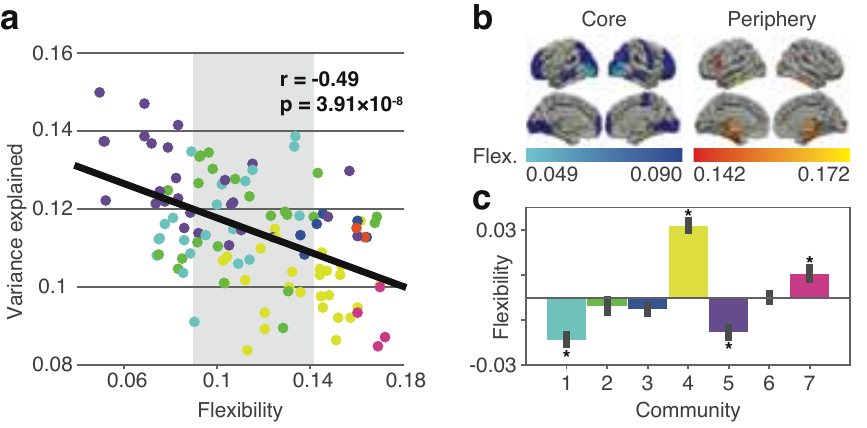}
\caption{\textbf{Regions in the network core are more associated with learning than regions in the network periphery.}
\emph{(a)} The variance in task accuracy explained by each node (calculated as the average variance explained across all edges departing from a node) is negatively correlated with node flexibility (Pearson's $r = -0.49, P<0.001$). Each circle corresponds to a brain region and is colored with the color of its corresponding community (Fig.~\ref{fig3}d). The shaded area corresponds to the 95\% confidence interval of the nodal null model described in (b).
\emph{(b)} A \emph{nodal null model} was constructed by rewiring the ends of the multilayer network's inter-layer edges uniformly at random 100 times. The temporal ``core'' was then defined as the set of regions whose mean nodal flexibility was below the $2.5\%$ confidence bound of the null-model distribution, and, similarly, the temporal ``periphery'' was defined as the set of regions whose mean nodal flexibility was above the $97.5\%$ confidence bound of the null-model distribution. The temporal core consists of regions of the visual, frontal, and (right) motor cortices. The temporal periphery consists of subcortical regions and regions of the anterior temporal lobe.
\emph{(c)} Average flexibility within each network community controlling for community size. Two communities exhibited flexibility significantly below that expected by its size: (i) fronto-parietal ($f=-0.019, P<0.001$), and (ii) visual ($f=-0.015, P<0.001$); and two communities exhibited flexibility significantly above that expected by its size: (i) fronto-temporal ($f=0.032, P<0.001$), and (ii) GP/NAcc ($f=0.011, P=0.019$).
\label{fig5}}
\end{figure}

We first examined the temporal variability of community structure by computing the \emph{flexibility} $f_i$ of each region $i$ \cite{bassett2013task}. To calculate flexibility, we identified the dynamic community structure over the 4 days of learning based on multilayer representations of temporal networks in 60 s windows \cite{mucha2010community}. We then calculated the flexibility of each region as the relative frequency with which it changed its allegiances to network communities over time: a high value of flexibility indicates that a region changes community affiliation frequently. We assessed the relationship between network flexibility and learning by calculating the amount of variance ($R$-squared) in task accuracy explained by each region: that is, the average $R$-squared over all edges departing from that region. We observed a strong negative relationship between node flexibility and variance explained: regions with low flexibility explained larger amounts of variance in task accuracy than regions with high flexibility (Pearson's $r = -0.49, P<0.001$; Fig.~\ref{fig5}a). 

These results suggest that the regions whose functional connectivity tracks task accuracy are those in a temporal core of relatively rigid areas whose affiliation with functional modules remains steady throughout task practice \cite{bassett2013task}. To determine whether this is indeed the case, we next categorized brain regions into \emph{temporal core} and \emph{temporal periphery} by assessing whether a region's flexibility was less than or greater than expected in a nodal null model, respectively (see Methods). Using this approach, we observed that the network core -- dynamically rigid regions with dense connectivity -- encompasses regions in the visual, frontal, and right motor areas. In contrast, the network periphery -- flexible regions with weak connectivity -- encompasses regions in the anterior temporal lobe and subcortical structures (Fig.~\ref{fig5}b). Moreover, the communities that we previously observed to be related to learning were not only the most rigid ones (Fig.~\ref{fig5}a,b), but were also less flexible than expected by their size alone (Fig.~\ref{fig5}c).

\subsection*{Predicting feedback condition from functional connectivity}

The results presented above characterize network features that change in concert with learning of value, broadly defined. Yet, information about stimulus value can be acquired in different ways, each involving potentially different brain structures \cite{delgado2012reward}. How might network recruitment and plasticity vary with the type of information used for learning? We addressed this question by comparing the functional network architecture during learning between the two feedback groups. 

The feedback manipulation introduced in the task design constrains the specific information available for the decision process while maintaining visual stimuli, participant's behaviors, and the overall task structure identical across conditions. The required mental operations, in turn, are expected to differ between groups despite identical visual inputs and motor outputs. Specifically, participants in the \emph{absolute} feedback group are required to retrieve and compare stimulus-specific value information, while participants in the \emph{relative} feedback group retrieve pairwise, relative relationships between stimuli based on previously reinforced choices. This is particularly expected early in the learning process, since after extensive training groups may converge to a state where both absolute and relative value information are available for the decision process. We thus hypothesized that task performance and learning may recruit different brain structures to the core value judgment network, in particular early in the learning process. In particular, we hypothesized a greater coupling of basal ganglia structures to output (motor) areas in participants of the \emph{relative} feedback group, with participants of the \emph{absolute} feedback group displaying stronger direct coupling between input (visual) and output (motor) structures.

We again used a data-driven classification procedure, identifying a different predictive network at each cross-validation fold and testing whether held-out data could be correctly labeled according to the type of feedback received by the participant. Specifically, we calculated the average functional connectivity for each participant in the first three scans (day 1). We then conducted a \emph{leave-two-out} cross-validation, forming a training set of all participants except for one participant in each group, and conducting a two-sample \emph{t}-test across subjects for each network edge. A predictive network for the absolute (relative) feedback group was defined as the set of edges with 1\% largest (smallest) \emph{t}-statistic, and the strength of the predictive network calculated for the two held-out participants.

\begin{figure}[!htbp]
\centering
\includegraphics[]{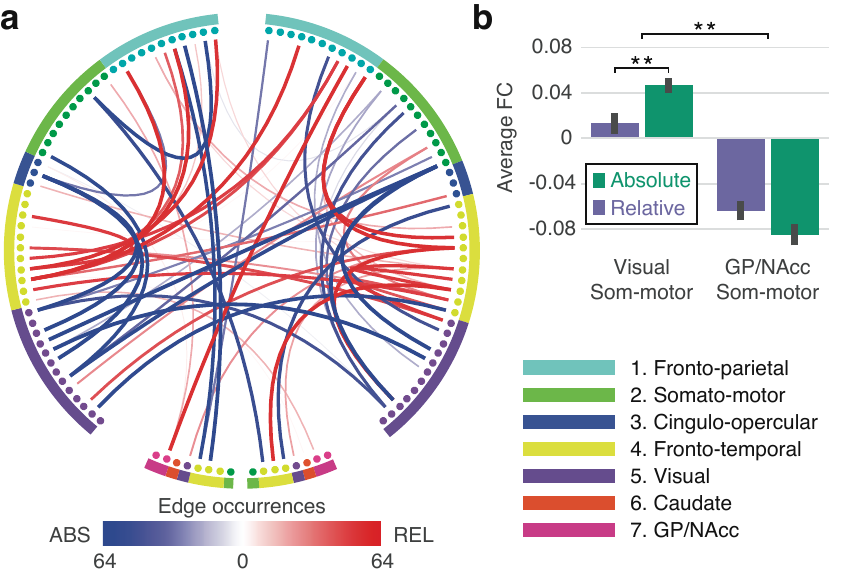}
\caption{\textbf{Functional connectivity between network modules predicts feedback condition.}
\emph{(a)} Using a cross-validation procedure, the subset of edges with lowest and highest \emph{t}-statistics in a two-sample \emph{t}-test between absolute and relative feedback were selected as the predictive network for the absolute and relative feedback groups, respectively. The figure shows the number of times each edge was part of the corresponding predictive networks (green: absolute feedback; purple: relative feedback). 
\emph{(b)} Average functional connectivity between Somato-motor and Visual, and between Somato-motor and GP/NAcc modules displayed separately for each feedback group. We observed a double dissociation: the visual module coupled more strongly to the somato-motor module in the \emph{absolute} feedback group, and the basal ganglia module composed of Nucleus Accumbens and Globus Pallidus coupled more strongly to the somato-motor module in the \emph{relative} feedback group, with the interaction being significant ($P=0.0027$).
\label{fig6}}
\end{figure}

We then asked, for each of the 64 cross-validation folds, which held-out participant had the largest network strength in each network. Using this procedure, we were able to label the held-out data with $79.69\%$ accuracy (chance: 50\%). To assess the significance of this result, we used a nonparametric permutation test in which the edges composing the predictive networks were selected uniformly at random. The observed classification accuracy in the true data was significantly greater than that observed in the null distribution ($P<0.0001$). The edges that appeared most frequently in the predictive network for \emph{absolute feedback} linked the visual community with the somato-motor, fronto-parietal and cingulo-opercular communities. In contrast, the edges that appeared most frequently in the predictive network for \emph{relative feedback} linked fronto-temporal areas to fronto-parietal and basal ganglia structures -- in particular the nucleus accumbens (Fig.~\ref{fig6}a). Similar results were obtained when using a Support-Vector Machine to classify feedback condition (accuracy: $80.49\%$; Fig. S5). In supplemental analyses at the community-level, we observed that feedback could be significantly classified based on interactions involving somato-motor, fronto-temporal, and caudate modules were modulated by feedback condition (Fig.S5).


Next, we examined the consistency of these effects across learning stages. We observed that classification accuracy decreased considerably when data from the final three scans (day 4) were used to define the predictive network and to classify held out data (accuracy: $68.75\%$, albeit still significantly greater than expected in the null distribution, as defined by a nonparametric permutation test: $P=0.0017$). These results suggest that the networks recruited for the two feedback conditions are distinct in the early stages of training, but become more similar to one another as subjects progressed in learning.

Finally, we explored whether the recruitment of systems hypothesized to be involved in each feedback condition would exhibit a double dissociation. We first subtracted the mean connectivity from each adjacency matrix in order to account for possible differences in overall network strength between conditions. We then calculated the average strength of connectivity between Somato-motor and Visual modules, and between Somato-motor and Basal Ganglia (Globus Pallidus and Nucleus Accumbens). In line with our hypotheses, we observed a significant interaction (two-way ANOVA interaction: $P=0.0027$), with connectivity between Somato-motor and Visual modules stronger for the \emph{absolute} feedback group (two-sample $t$-test on Fisher normalized correlation values: $t(14) = 3.02, P=0.0092$), and connectivity between the Somato-motor and Basal Ganglia modules stronger (though not significantly) for the \emph{relative} feedback group (two-sample $t$-test on Fisher normalized correlation values: $t(14) = 1.72, P=0.11$; Fig.~\ref{fig6}b). No differences were significant on day 4 (two-sample $t$-tests: $t(14) = 0.30, P=0.77$, $t(14) = 0.07, P=0.95$; two-way ANOVA interaction: $P=0.80$). Together, these results indicate a differential recruitment of brain structures for the two feedback conditions, especially early on, with output areas (e.g. somato-motor) more connected to visual areas or to basal ganglia structures in absolute and relative feedback conditions, respectively. 

\clearpage
\section*{Discussion}

In this study, we investigate the network-level neural markers of human value learning. Using neuroimaging data collected from a cohort of 20 healthy adults as they learned the monetary value of 12 novel visual stimuli, we demonstrate that functional connectivity -- patterns of statistical dependencies between activity in different brain regions -- varies in a manner that directly tracks changes in behavior. As accuracy increases, functional connectivity within areas of the visual system and between visual and fronto-cingulate regions also increases. This pattern of changes is consistent with a novel theory of human learning that posits that changes in behavior are driven by changes in the brain's dynamic network core, a set of regions that maintain their affiliation to functional modules (rigidity) while altering the specific pattern of functional connections (dynamics). Intriguingly, these patterns of reconfiguration can be used to predict the learning stage of a single subject, and also to predict which type of learning feedback a subject has been exposed to. Collectively, our results provide strong evidence for the notion that functional brain networks are sensitive to behavioral improvements and task conditions, and offer novel insights into the cognitive neuroscience of learning.

\paragraph{Task-based network architecture}

Our study begins with a demonstration that functional networks defined during task execution are sensitive to learning, whereas resting state networks are not (Fig.~\ref{fig2}a). While the network architectures of rest and task states certainly display some commonalities \cite{cole2014intrinsic}, the present work provides evidence for their differences, complementing a growing set of studies particularly in the context of learning \cite{bassett2013task,bassett2015learning}. Indeed, the community structure observed in our study differs from that observed in the resting state \cite{power2011functional}, being characterized by modules for visual perception, value comparison, decision-making, and motor responses, as well as a module in which the fronto-parietal and its resting competitor (the default mode) are strongly intertwined \cite{fornito2012competitive}. Interestingly, the cingulo-opercular network, thought to subserve goal directed behavior through the stable maintenance of task sets\cite{dosenbach2007distinct}, played an important role in the present task. This is likely a consequence of it being predominantly composed of the anterior cingulate, an area involved in reward processing and decision making \cite{kennerley2006optimal}. Indeed, this diversity of functions has been discussed in depth in recent reviews \cite{shenhav2013expected,holroyd2012motivation}. Finally, the separation of subcortical structures in the basal ganglia into isolated communities (caudate and nucleus accumbens with globus pallidus) is not characteristic of resting state dynamics, and indicates that these structures have a timecourse of neural responses that are distinct from other communities, potentially due to signals related to prediction errors \cite{o2003temporal}. 

\paragraph{A core-periphery model of value learning} 

Recent studies have started to explore the important question of how different cognitive systems interact with one another during task execution, with several taking an explicit view of cognitive systems as network modules whose composition may change over time \cite{mattar2015functional,braun2015dynamic,bassett2015learning}. Using a similar approach, we demonstrate that the modules that we observe are composed of brain regions that differ in their roles in the larger dynamic network \cite{mattar2015functional}. As the subjects learn, functional connections (or edges within the network) change in their strength collectively enabling a reconfiguration of the network of interactions between brain areas, potentially tracking changes in shared information or communication \cite{friston2011functional}. In the context of a value judgment task, our data demonstrate that these changes can be parsimoniously described by a core-periphery model of brain network dynamics \cite{bassett2013task}: a temporally stable and densely connected set of core regions, as well as a temporally flexible and sparsely connected set of peripheral regions, can be singled out from the remaining bulk of the network. Prior evidence suggests that this specific type of network dynamics enables a careful balance between the temporal consistency of functional connections required for task execution, and the temporal flexibility of functional connections required for transient control of behavior \cite{bassett2013task}. Such observations at the mesoscale level constitute some of the greatest strengths of a network-based approach, which is able to identify emergent properties at a broader level than specific connections.

\paragraph{A generalized theory of network reconfiguration during learning}

The core-periphery model of value learning lays the foundation for a generalized theory of network reconfiguration during learning. We posit that learning requires a change in the functional integration between modules located in the temporal core of a dynamic network. This theory is built on prior work suggesting that modules in the temporal core may be specialized for the task at hand while modules in the temporal periphery may be recruited transiently for more domain-general processes \cite{fedorenko2014reworking}. Moreover, the theory draws on prior evidence for a core-periphery network dynamic in a different type of learning \cite{bassett2013task}, where change in behavior is associated with changes in the strength of functional connections between modules located in the temporal core \cite{bassett2015learning}. Our results provide key support for this theory: we observe that regions in the network core, or those with relatively low flexibility, tend to be the most relevant to learning, while regions in the network periphery characterized by fleeting community allegiance do not track behavioral change. 

However, we note that within the context of this broader paradigm, the inter-module adaptations characteristic of a learning task are likely specific to the material learned. During motor skill learning, for example, motor and visual systems that make up the temporal core \cite{bassett2013task} transition from being heavily integrated early in learning to functioning as relatively autonomous units after extensive practice \cite{bassett2015learning}. This growing autonomy indicates that motor and visual regions exhibit similar temporal profiles of BOLD activity early in learning, but distinct temporal profiles of BOLD activity later in learning. In contrast, our results indicate that value learning is characterized by a growing integration (not autonomy) between the visual and fronto-cingulate modules that make up the temporal core. The processes of growing integration \emph{versus} autonomy are likely a consequence of distinct cognitive demands in the two tasks. During visuo-motor sequence learning, a person initially relies on a visual cue to perform a finger movement, an action that requires integration between motor and visual cortices; however, once a sequence becomes overlearned, a subject has mastered direct motor-motor associations where a given finger movement is the cue for the next finger movement, enabling autonomy of the visual system. In contrast, a task involving decision-making cannot be solved by motor-motor associations, but instead must rely on an increased synchrony between areas that encode and process sensory and value information to allow for swifter and more accurate responses. 

\paragraph{Integration of visual and valuation systems}

Models and theory aside, it is important to couch our network-level observations within the cognitive neuroscience of value. Our task paradigm requires participants to process incoming visual information, recognize the stimulus pair, and retrieve the relevant value information, before making a choice and entering a response. With the feedback received, the stored values of the recently observed stimuli are updated so that future choices are optimized for accuracy. Conceptually speaking, both at the time of retrieval and at the time of updating, visual and value-related information must be combined, requiring an intricate interplay between the visual system \cite{van1992information} and the valuation system \cite{bartra2013valuation}. Our results directly confirm these predictions by demonstrating that visual areas, and in particular their interactions with areas of the cingulo-opercular module, are strongly modulated by learning, becoming increasingly connected as task accuracy increases (Fig.~\ref{fig3}e). The increased connectivity between these two modules means that their activity becomes increasingly synchronized with learning. A plausible explanation for this finding is that activity in visual regions become increasingly modulated by value -- e.g. responding more strongly when currently attended stimuli have higher values --, similarly to what is observed in regions of the valuation system \cite{bartra2013valuation}. Indeed, it has recently been observed that visual responses after value learning become modulated by value similarity \cite{persichetti2015value}. While our results do not explicitly test for the emergence of multivariate representations that are sensitive to value similarity, they are in direct agreement with this possibility.

\paragraph{Effect of feedback}

Even in a task as simple as a two-alternative forced choice, the selection of the alternative with highest value may require widely different processes depending on the information previously encoded and currently available. In our paradigm, subjects in the \emph{relative} feedback group learned ordinal information while those in the \emph{absolute} feedback group learned cardinal information, and so the content learned itself differed between conditions. While most of the research on feedback-based learning has been successfully formalized within a reinforcement learning framework, where a positive (or negative) prediction error signal reinforces (weakens) specific stimulus-response associations, this framework cannot account for all forms of learning. In particular, learning of declarative information (in our task also received through feedback), linking a stimulus with a monetary value, is likely to require a different strategy given the lack of a clear reinforcement signal \cite{squire1992declarative,packard2002learning}. These two different learning strategies are thought to recruit distinct memory systems, with basal ganglia regions mediating learning through trial and error (right vs. wrong), and more posterior regions (in particular the medial-temporal lobe) mediating declarative learning \cite{squire1992declarative,packard2002learning,delgado2012reward}.

Recent evidence suggest that the relationship between these two systems is cooperative rather than competitive \cite{voermans2004interaction,cincotta2007dissociation,dickerson2011parallel}. In particular, some authors suggest that sensory information may pass through both trial and error and declarative memory systems independently \cite{white2002multiple} before reaching output, motor systems. Our results provide direct support to theories of multiple learning systems by demonstrating that feedback condition can be robustly classified based on functional networks. Furthermore, in line with an cooperative interaction view, we show that the output somato-motor module couples differentially with basal ganglia structures or with a visual module depending on feedback type (Fig.~\ref{fig6}b).

\paragraph{Methodological considerations}

It has become increasingly clear that the insights obtained with a network-based approach are complementary and, for the most part distinct from those obtained with univariate or multivariate activation approaches \cite{bassett2015learning,siebenhuhner2013}. Activation-based approaches typically rely on the assumption that the regions that are relevant for a cognitive process responds with different average intensities during that process in comparison to a baseline \cite{friston2005}. In contrast, the assumptions in network based approaches are that the pattern of statistical dependencies \emph{between} regions is what varies between conditions, not requiring an overall bulk change in average activity nor the definition of a baseline to which all conditions are compared to \cite{bullmore2011brain}. While activation-based studies have provided us with many important and meaningful insights on behavioral and cognitive neuroscience over the course of the last two decades, network studies in task contexts have flourished more slowly and only recently the tools, statistics and diagnostic approaches necessary for their analyses and interpretation have started to be developed \cite{medaglia2015cognitive}. 

Our study offers important methodological advancements for the network neuroscientist's toolkit. First and foremost, a major contribution of our study is the direct relationship between variables derived from behavior and modulations in functional connectivity. A key advantage of this technique over more conventional neuroimaging approaches is the ability to infer the specific functional interactions in the brain that are associated with changes in behavior \cite{shehzad2014}. Secondly, our work is also part of a growing set of studies that use completely data-driven analysis approaches to identify network components relevant for a cognitive process, and a cross-validated procedure for out-of-sample prediction, which may reduce the risk of overfitting and improve generalizability \cite{turk2013functional, shirer2012decoding, rosenberg2015neuromarker, finn2015functional}.

\paragraph{Implications for cognitive and clinical neuroscience}

The learning of value is a fundamental prerequisite for healthy adult human behavior. A natural question that arises is how the network architecture of value learning reconfigures as children develop from infants through adolescence and into adulthood, a process that is known to be accompanied by changes in both structural \cite{betzel2014changes} and functional \cite{gu2015emergence} brain networks. Moreover, our results motivate the hypothesis that the normative characteristics of network reconfiguration that we observe in this study will be fundamentally altered in patients with deficits in reinforcement learning behavior, and in fronto-cingulate cognitive control systems, including individuals with Parkinson's disease or schizophrenia. Finally, it could be interesting in future studies to assess neurotransmitter-level drivers of these reconfiguration dynamics, and their dependence on cellular-level mechanism of synpatic plasticity.

\newpage
\section*{Methods}
\label{sec:methods}

\subsection*{Participants}

Twenty participants (nine female; ages 19--53 years; mean age = 26.7 years) with normal or corrected vision and no history of neurological disease or psychiatric disorders were recruited for this experiment. All participants volunteered and provided informed consent in writing in accordance with the guidelines of the Institutional Review Board of the University of Pennsylvania (IRB \#801929).

\subsection*{Experimental setup and procedure}

Participants learned the monetary value of 12 novel visual stimuli in a reinforcement learning paradigm. Learning occurred over the course of four MRI scan sessions conducted on four consecutive days.

The novel stimuli were 3-dimensional shapes generated with a custom built MATLAB toolbox (code available at \href{http://github.com/saarela/ShapeToolbox}{http://github.com/saarela/ShapeToolbox}) and rendered with RADIANCE \cite{ward1994radiance}. ShapeToolbox allows the generation of three-dimensional radial frequency patterns by modulating basis shapes, such as spheres, with an arbitrary combination of sinusoidal modulations in different frequencies, phases, amplitudes, and orientations. A large number of shapes were generated by selecting combinations of parameters at random. From this set, we selected twelve that were considered to be sufficiently distinct from one another. A different monetary value, varying from \$1.00 to \$12.00 in integer steps, was assigned to each shape. These values were uncorrelated with any parameter of the sinusoidal modulations, so that visual features were not informative of value.

On each trial of the experiment, participants were presented with two shapes side by side on the screen and asked to choose the shape with the higher monetary value in an effort to maximize the total amount of money in their bank. The shape values on a given trial were independently drawn from a Gaussian distribution with mean equal to the true monetary value and the standard deviation equal to \$0.50. This variation in the trial-specific value of a shape was incorporated in order to ensure that participants thought about the shapes as having worth, as opposed to simply associating a number or label with each shape. 

Participants completed 20 minutes of the main task protocol on each scan session, learning the values of the 12 shapes through feedback. The sessions comprised of three scans of 6.6 minutes each, starting with 16.5 seconds of a blank gray screen, followed by 132 experimental trials (2.75 seconds each), and ending with another period of 16.5 seconds of a blank gray screen. Stimuli were back-projected onto a screen viewed by the participant through a mirror mounted on the head coil and subtended 4 degrees of visual angle, with 10 degrees separating the center of the two shapes. Each presentation lasted 2.5 seconds and, at any point within a trial, participants entered their responses on a 4-button response pad indicating their shape selection with a leftmost or rightmost button press. Stimuli were presented in a pseudorandom sequence with every pair of shapes presented once per scan.

Feedback was provided as soon as a response was entered and lasted until the end of the stimulus presentation period. Participants were randomly assigned to two groups depending on the type of feedback received. In the RELATIVE feedback case, the selected shape was highlighted with a green or red square, indicating whether the selected shape was the most valuable of the pair or not, respectively. In the ABSOLUTE feedback case, the actual value of the selected shape (with variation) was displayed in white font. Importantly, no information about the correctness of the choice was given in the ABSOLUTE feedback case. Between each run, both groups received feedback about the total amount of money accumulated up to that point.

In addition to the main learning protocol, we collected fMRI data during a functional localizer, two scans of a size judgment task, and one scan of a value judgment task. The functional localizer scans consisted of 10 s blocks of faces, scenes, objects and scrambled objects (800 ms presentation and 200 ms inter-stimulus interval) as participants performed a one-back task on image repetition. The size judgment task scans consisted of consecutive presentations of shapes drawn from the set and presented with a $\pm$ 10\% size modulation (1500 ms presentation and 250 ms inter-stimulus interval) as participants indicated whether the shape was presented in a slightly larger or smaller variation. The value judgment task scans consisted of consecutive presentations of shapes drawn from the set (1500 ms presentation and 250 ms inter-stimulus interval) as participants indicated whether the shape was one of the six least or one of the six most valuable shapes. No feedback was given in any of these tasks. Data from these additional scans were not included in any of the present analyses.

\subsection*{Subject exclusion criteria}

Participants were excluded on the basis of head motion and task performance. From the set of 20 participants recruited for the experiment, two were excluded due to excessive head motion (criterion: average absolute or relative motion greater than two standard deviations away from the mean, Fig SX), and two were excluded due to low task performance (criterion: average task accuracy at the final scan lower than 80\%, Fig SX). Our investigation therefore included 16 participants (eight female; ages 19-31 years; mean age = 24.1 years), of which eight remained in each feedback condition. This sample size is consistent with accepted good practices in this field \cite{desmond2002estimating}.

\subsection*{MRI Data collection}

Magnetic resonance images were obtained at the Hospital of the University of Pennsylvania using a 3.0 T Siemens Trio MRI scanner equipped with a 32-channel head coil. T1-weighted structural images of the whole brain were acquired on the first scan session using a three-dimensional magnetization-prepared rapid acquisition gradient echo pulse sequence (repetition time (TR) 1620 ms; echo time (TE) 3.09 ms; inversion time 950 ms; voxel size 1 mm $\times$ 1 mm $\times$ 1 mm; matrix size 190 $\times$ 263 $\times$ 165). A field map was also acquired at each scan session (TR 1200 ms; TE1 4.06 ms; TE2 6.52 ms; flip angle 60\degree; voxel size 3.4 mm $\times$ 3.4 mm $\times$ 4.0 mm; field of view 220 mm; matrix size 64 $\times$ 64 $\times$ 52) to correct geometric distortion caused by magnetic field inhomogeneity. In all experimental runs with a behavioral task, T2*-weighted images sensitive to blood oxygenation level-dependent contrasts were acquired using a slice accelerated multiband echo planar pulse sequence (TR 2,000 ms; TE 25 ms; flip angle 60\degree; voxel size 1.5 mm $\times$ 1.5 mm $\times$ 1.5 mm; field of view 192 mm; matrix size 128 $\times$ 128 $\times$ 80). In all resting state runs, T2*-weighted images sensitive to blood oxygenation level-dependent contrasts were acquired using a slice accelerated multiband echo planar pulse sequence (TR 500 ms; TE 30 ms; flip angle 30\degree; voxel size 3.0 mm $\times$ 3.0 mm $\times$ 3.0 mm; field of view 192 mm; matrix size 64 $\times$ 64 $\times$ 48).

\subsection*{MRI data preprocessing}

Cortical reconstruction and volumetric segmentation of the structural data was performed with the Freesurfer image analysis suite \cite{dale1999cortical}. Boundary-Based Registration between structural and mean functional image was performed with Freesurfer \emph{bbregister} \cite{greve2009accurate}.

Preprocessing of the resting state fMRI data was carried out using FEAT (FMRI Expert Analysis Tool) Version 6.00, part of FSL (FMRIB's Software Library, \href{http://www.fmrib.ox.ac.uk/fsl}{www.fmrib.ox.ac.uk/fsl}). The following pre-statistics processing was applied: EPI distortion correction using FUGUE \cite{jenkinson2004improving}; motion correction using MCFLIRT \cite{jenkinson2002improved}; slice-timing correction using Fourier-space time series phase-shifting; non-brain removal using BET \cite{smith2002fast}; grand-mean intensity normalization of the entire 4D dataset by a single multiplicative factor; highpass temporal filtering (Gaussian-weighted least-squares straight line fitting, with sigma=50.0s).

Nuisance time series were voxelwise regressed from the preprocessed data. Nuisance regressors included (i) three translation ($X, Y, Z$) and three rotation ($pitch, yaw, roll$) time series derived by retrospective head motion correction ($R=[X,Y,Z,pitch,yaw,roll]$), together with expansion terms ([$R$,$R^2$,$R_{t-1}$,$R_{t-1}^2$]), for a total of 24 motion regressors \cite{friston1996movement}); (ii) the five first principal components of non-neural sources of noise, estimated by averaging signals within white matter and cerebrospinal fluid masks, obtained with Freesurfer segmentation tools and removed using the anatomical CompCor method (aCompCor) \cite{behzadi2007component}; and (iii) an estimate of a local source of noise, estimated by averaging signals derived from the white matter region located within a 15 mm radius from each voxel, using the ANATICOR method \cite{jo2010mapping}. Global signal was not regressed out of voxel time series due to its controversial application to resting state fMRI data \cite{murphy2009impact, saad2012trouble, chai2012anticorrelations}.

\subsection*{Network construction}

Network analyses of brain function began with a definition of the interacting units -- nodes -- and a quantification of the interaction -- edges. Our study follows standard practices in the field of neuroimaging and defines nodes as a collection of contiguous voxels given by an atlas or parcellation scheme, and further defines edges as the statistical dependence between the average activity in the corresponding nodes. Consistent with previous studies of task-based functional connectivity during learning, we parcellated the brain into 112 cortical and subcortical regions, separated by hemisphere using the structural Harvard-Oxford atlas of the FMRIB (Oxford Centre for Functional Magnetic Resonance Imaging of the Brain) Software Library (FSL; Version 5.0.4). We warped the MNI152 regions into subject-specific native space using FSL FNIRT and nearest-neighbor interpolation and calculated the average BOLD signal across all gray matter voxels within each region. The participant's gray matter voxels were defined using the anatomical segmentation provided by Freesurfer, projected into subject's EPI space with  \emph{bbregister}.

We then calculated the edge weights connecting nodes by measuring the wavelet coherence between the activities of the corresponding regions. We first extracted the wavelet coefficients of the average signal within each region using the WMTSA Wavelet Toolkit for MATLAB \cite{percival2006wavelet}. Given our sampling frequency of 2 s, we used scale 2 coefficients (corresponding approximately to $0.06--0.125$ Hz) to calculate the magnitude squared coherence, using the MATLAB function \textit{mscohere}. We repeated this procedure for all pairs of regions, forming an adjacency matrix $\textbf{A}$ for each subject and for each scan.

\subsection*{Multislice Community Detection}

Approaches to data clustering in networks are generally based on community detection techniques \cite{porter2009communities,fortunato2010community}. Here we use a generalized Louvain method \cite{blondel2008fast} for optimizing modularity \cite{newman2006modularity} developed specifically for community detection in multislice systems \cite{mucha2010community}: systems in which multiple networks linked by an ordered or categorical dimension (time in our case) are to be examined at once. We implemented a categorical multislice modularity maximization \cite{mucha2010community,jutla2011generalized} which considers the multiple adjacency matrices as slices of a single network, enforcing consistency in node identity across slices by adding interslice connections between each node and itself across adjacent slices of the network. We then optimize the multislice modularity quality function \eqref{eq:modularity}, which uses the relative densities of intra-community connections versus inter-community connections to identify a partition of network nodes into communities or modules \cite{mucha2010community}, defined as:

\begin{equation} \label{eq:modularity}
	Q_{\text{multislice}} = \frac{1}{2 \mu} \sum_{ijsr}
		\left[
			\left(
				A_{ijs} - \gamma_s V_{ijs}
			\right) \delta_{sr} + \delta_{ij} \omega_{jsr}
		\right] \delta (g_{is} , g_{jr})
\end{equation}
where the adjacency matrix of slice $s$ has components $A_{ijs}$, the element $V_{ijs}$ gives the components of the corresponding matrix for a null model, $\gamma_s$ is the structural resolution parameter of slice $s$, the quantity $g_{is}$ gives the community (i.e., ``module'') assignment of node $i$ in slice $s$, the quantity $g_{jr}$ gives the community assignment of node $j$ in slice $r$, the parameter $\omega_{jsr}$ is the connection strength between node $j$ in slice $s$ and node $j$ in slice $r$, the total edge weight in the network is $\mu = \frac{1}{2}\sum_{jr}\kappa_{jr}$, the strength of node $j$ in slice $s$ is $\kappa_{js} = k_{js} + c_{js}$, the intraslice strength of node $j$ in slice $s$ is $k_js$, and the interslice strength of node $j$ in slice $s$ is $c_{js} = \sum_r \omega_{jsr}$. We employ the Newman-Girvan null model within each layer by using

\begin{equation} \label{eq:newman}
	V_{ijs} = \frac{k_{is} k_{js}}{2 m_s}
\end{equation}
where $m_s = \frac{1}{2}\sum_{ij} A_{ijs}$ is the total edge weight in slice $s$. The free-parameters are the structural resolution parameters, $\gamma_s$, and the interslice coupling parameters, $\omega_{jsr}$, here assumed to be constant ($\gamma_s = \gamma, \forall s$ and $\omega_{jsr} = \omega, \forall j$ and $\forall s \neq r$, meaning that node $j$ in slice $s$ connects to node $j$ in every slice $r \neq s$ with weight $\omega$). These parameters control the size of communities within a given layer and the number of communities discovered across layers, respectively. 

In order to obtain a single, representative, partition of the brain into network communities, we performed 100 opimizations of the modularity function for each participant, using non-overlapping 60 s windows and the standard parameters of $\gamma=1.0$ and $\omega=1.0$. We then calculated the \emph{module allegiance} matrix \cite{mattar2015functional}, a $112 \times 112$ matrix whose $i,j$ elements corresponds to the probability that regions $i$ and $j$ belong to the same community across all optimizations, scans, and participants. By repeating the procedure of maximizing the modularity function on the module allegiance matrix 100 times and re-calculating a new module allegiance matrix, a consensus partitition is considered to have been obtained when all 100 optimizations of the modularity function are identical. The structural resolution parameter used in the opimizations of the single-layer module allegiance matrix can be tuned to yield a different level of the hierarchical organization of the network. We chose a value of $\gamma = 1.4$, which yielded the maximum number of communities present simultaneously in both hemispheres of the brain.

\subsection*{Flexibility and core-periphery structure}

As in previous work \cite{bassett2011dynamic,bassett2013task}, we define flexibility of a brain region as the proportion of task conditions in which this region changes its community allegiance. In order to determine the significance of a brain region's flexibility, we contrasted its value with a set of values obtained from a \emph{nodal null mode}. We rewired the ends of the multilayer network's inter-layer edges (which connect nodes in one layer to nodes in another) uniformly at random. After applying the associated permutation, an inter-layer edge can, for example, connect node $i$ in layer $t$ with node $j \neq i$ in layer $t+1$ rather than being constrained to connect each node $i$ in layer $t$ with itself in layer $t+1$.

We performed 100 different rewirings to construct a set of 100 nodal null-models for each subject. We then averaged these results across subjects, yielding a distribution of flexibility values that would be expected if an edge identity was not preserved over time. By comparing the true flexibility value of a region with its empirical null distribution, we defined the temporal ``core'' of the network as the set of regions whose flexibility values are below the 2.5\% confidence bound of the null distribution, the temporal ``periphery'' of the network as the set of regions whose flexibility values are above the 97.5\% confidence bound of the null distribution, and the temporal ``bulk'' as the set of remaining nodes (i.e. confined to the 95\% confidence interval of the null distribution). A more more in-depth description of this approach and its applications to the study of human skill-learning is provided elsewhere \cite{bassett2013task}.

\clearpage
\addcontentsline{toc}{chapter}{Bibliography}
\bibliographystyle{ieeetr}
\bibliography{bibfile.bib}

\clearpage
\section*{Supporting Information}
\label{sec:supplement}
\beginsupplement

\/* 

\textbf{S1 Text. Supplementary Methods.}
\\
\textbf{S2 Text. Supplementary Results.} 
\\
\textbf{S3 Text. Supplementary Discussion.} 
\\
\textbf{S1 Table. Abbreviations.} List of abbreviations for cognitive systems 
\\

*/ 

\begin{figure}[!htbp]
\centering
\includegraphics{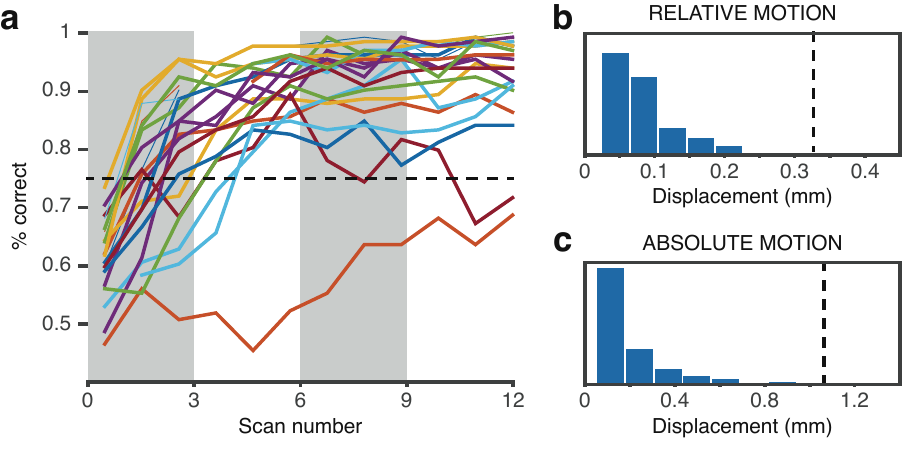}
\caption{\textit{Related to Figure 1}. \textbf{Subject exclusion criteria.}
\\*(a) Individual learning curves for each of the 20 participants. Participants with task performance below 75\% at the last run were excluded from the analyses in the main text.
\\*(b) Average relative motion within scan runs. Participants with average relative motion larger than three standard deviations away from the mean in two of more scans were excluded from the analyses in the main text. 
\\*(c) Average absolute motion within scan runs. Participants with average relative motion larger than three standard deviations away from the mean in two of more scans were excluded from the analyses in the main text.
\label{FigureS1}}
\end{figure}

\newpage
\begin{figure}[!htbp]
\centering
\includegraphics{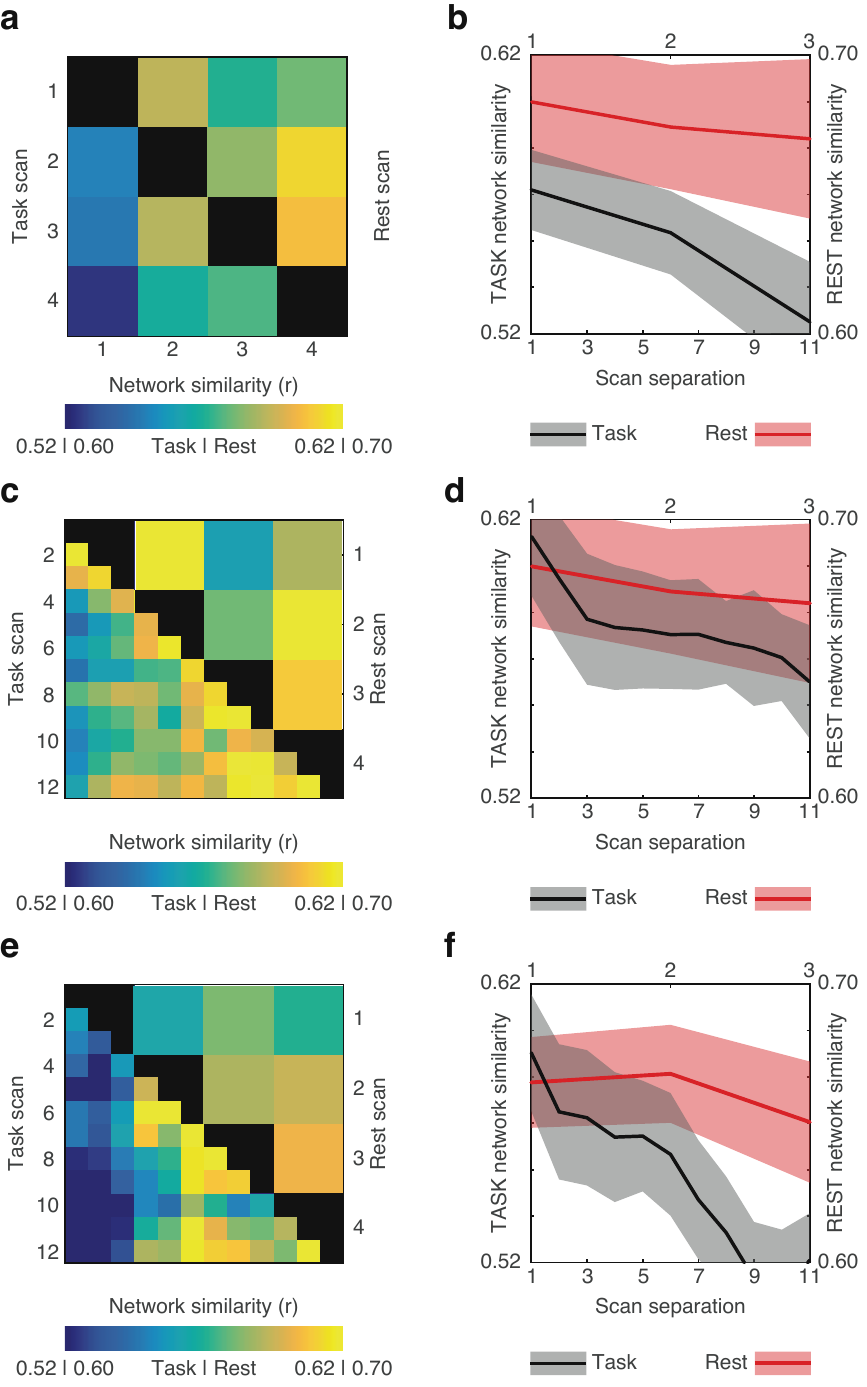}
\caption{\textit{Related to Figure 2}. \textbf{Network evolution throughout learning.}
\\*(a) Network similarity limited to first task scans on each day.
\\*(b) Corresponding average network similarity as a function of temporal separation between corresponding scans. Black line: task scans; Red line: rest scans.
\\*(c) Network similarity for participants from \textit{absolute} feedback group.
\\*(d) Corresponding average network similarity as a function of temporal separation between corresponding scans. Black line: task scans; Red line: rest scans.
\\*(e) Network similarity for participants from \textit{relative} feedback group.
\\*(f) Corresponding average network similarity as a function of temporal separation between corresponding scans. Black line: task scans; Red line: rest scans.
\label{FigureS2}}
\end{figure}

\newpage
\begin{figure}[!htbp]
\centering
\includegraphics{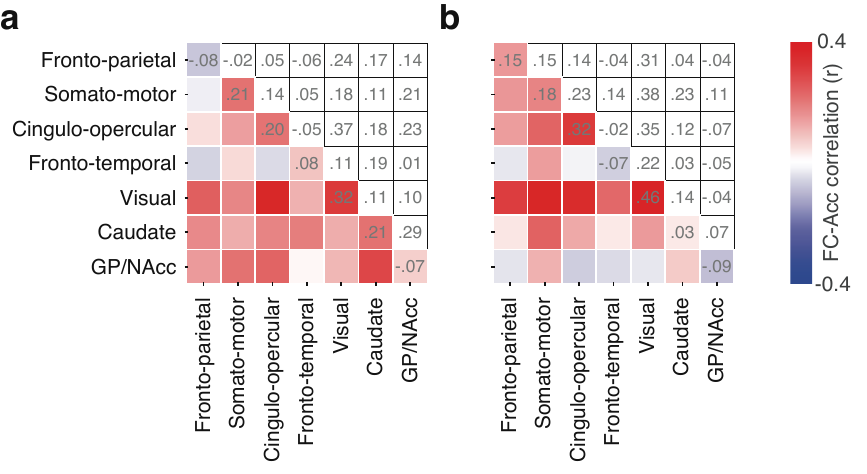}
\caption{\textit{Related to Figure 3}. \textbf{Community-level interactions related to value learning}
\\*(a) Correlation between average edge weight within/between communities and task accuracy for \emph{absolute} feedback group.
\\*(b) Correlation between average edge weight within/between communities and task accuracy for \emph{relative} feedback group.
\label{FigureS3}}
\end{figure}

\newpage
\begin{figure}[!htbp]
\centering
\includegraphics{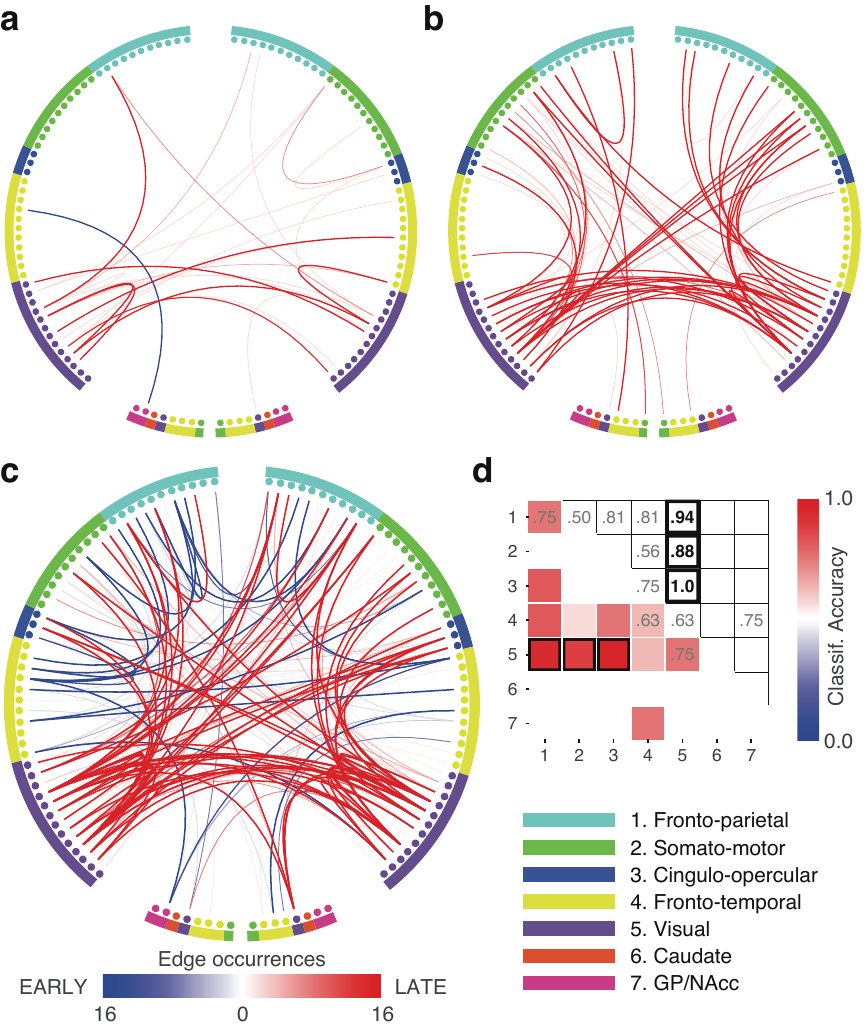}
\caption{\textit{Related to Figure 4}. \textbf{Prediction of learning stage from functional networks}
\\*(a) Related to Fig. 4a but limited to \emph{absolute} feedback group. Using a cross-validation procedure, the subset of edges where a one-sample $t$-test yielded a test statistic with $P$-value lower than $0.001$ was selected as a \emph{predictive network}. The figure shows the number of cross-validation folds in which each edge was identified as part of the predictive network. Edges whose strength correlated positively with task accuracy are displayed in red. Edges whose strength correlated negatively with task accuracy are displayed in blue. 
\\*(b) Related to Fig. 4a but limited to \emph{relative} feedback group. 
\\*(c) Related to Fig. 4a (data from both groups), but forming predictive networks after the removal of the average network strength at each scan. Using this alternative cross-validation procedure, the number of edges present in the predictive networks ranged from 105 to 154 ($M = 132.5, SD = 13.7$) and classification was correct in all 16 participants (accuracy: $100\%$ vs. chance: $50\%$; one-tailed binomial test: $P=1.525 \times 10^{-5}$).
\\*(d) Related to Fig. 4b (data from both groups), but forming predictive networks after the removal of the average network strength at each scan. Cell colors and numbers represent classification accuracy in labeling held-out data as coming from scans early or late in the learning process. Data from the left-out participant was significantly classified above chance (50\%) when the predictive network was comprised of edges connecting: (i) visual and fronto-parietal modules (accuracy: $93.75\%$; one-tailed binomial test, adjusted $P$-value: $P=0.0054$); (ii) visual and somato-motor modules (accuracy: $87.50\%$; one-tailed binomial test, adjusted $P$-value:  $P=0.044$); and (iii)  visual and cingulo-opercular modules (accuracy: $100\%$; one-tailed binomial test, adjusted $P$-value:  $P=0.00032$).
\label{FigureS4}}
\end{figure}

\newpage
\begin{figure}[!htbp]
\centering
\includegraphics{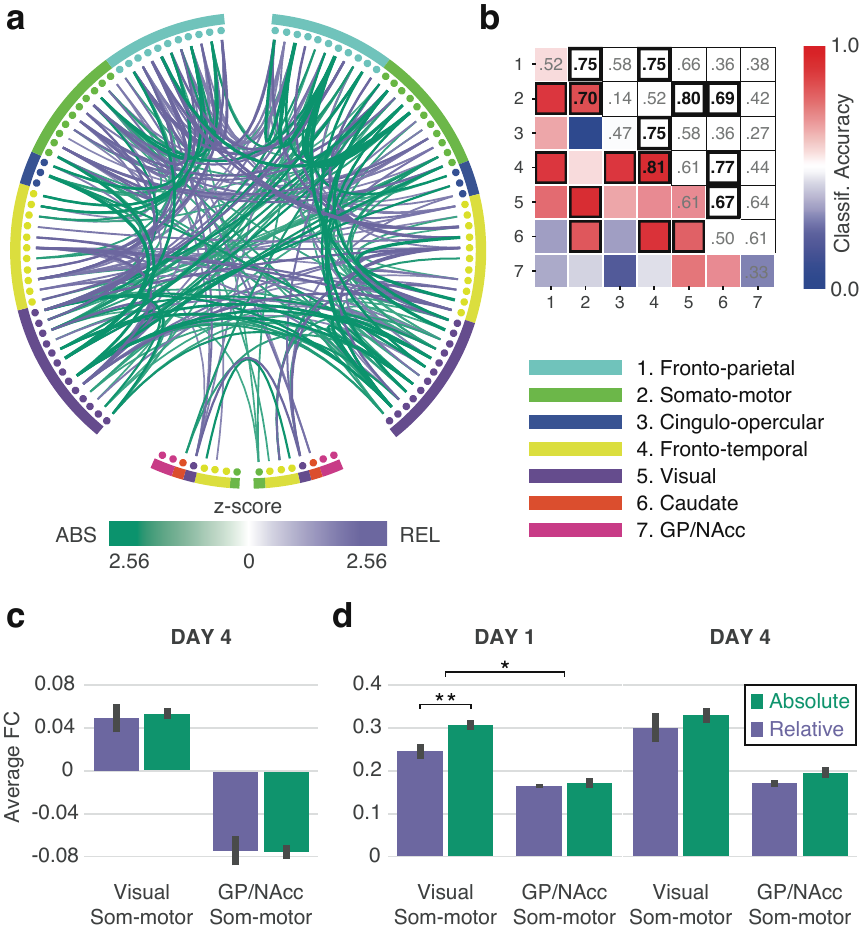}
\caption{\textit{Related to Figure 6}. \textbf{Prediction of feedback-type from functional networks}
\\*(a) We used a support-vector machine with leave-two-out cross-validation to classify feedback type based on the entire set of edge weights. On each cross-validation fold, the $w$-map represents how useful each feature (edge weight) is at discriminating between conditions. The figure displays the average z-scored $w$-map, limited to connections with $z > 1.96$ (green: absolute feedback; purple: relative feedback). Notice the similarity with Fig. 6a.
\\*(b) To gain insight into the specific modules that enable classification of feedback-type, we conducted the analyses in the main text separately for each pair of communities, selecting the top 10\% edges at each cross-validation fold. Cell colors and numbers represent classification accuracy in labeling held-out data as coming from participants in the absolute-vs.-relative feedback group. Communities whose interactions classified held-out data significantly (permutation tests, Bonferroni corrected at $\alpha=0.05$) are highlighted. Community order is displayed in the bottom-right.We observed that interactions involving somato-motor, fronto-temporal, and caudate modules were modulated by feedback type.
\\*(c) Related to Fig. 6b. Average functional connectivity in \textbf{DAY 4} between Somato-motor and Visual, and between Somato-motor and GP/NAcc modules displayed separately for each feedback group. Neither difference nor the interaction was significant (two-sample $t$-tests: $t(14) = 0.30, P=0.77$, $t(14) = 0.07, P=0.95$; two-way ANOVA interaction: $P=0.80$).
\\*(d) Related to Fig. 6b. Average functional connectivity in \textbf{DAY 1} (left) and \textbf{DAY 4} (right) between Somato-motor and Visual, and between Somato-motor and GP/NAcc modules displayed separately for each feedback group. Similarly to the results presented in the main text, we observed that connectivity between Somato-motor and Visual modules was stronger for the \emph{absolute} feedback group (two-sample $t$-test on Fisher normalized correlation values: $t(14) = 3.14, P=0.0073$), while connectivity between the Somato-motor and Basal Ganglia modules was not significantly different between groups (two-sample $t$-test on Fisher normalized correlation values: $t(14) = 0.51, P=0.62$), though the interaction term was significant (two-way ANOVA interaction: $P=0.022$; Fig.~\ref{fig6}b). No differences were significant on day 4 (two-sample $t$-tests: $t(14) = 0.86, P=0.40$, $t(14) = 1.82, P=0.09$; two-way ANOVA interaction: $P=0.88$).
\label{FigureS5}}
\end{figure}

\clearpage
\section*{Acknowledgements}
DSB and MGM acknowledge support from the John D. and Catherine T. MacArthur Foundation, the Alfred P. Sloan Foundation, the Army Research Laboratory through contract no. W911NF-10- 2-0022 from the U.S. Army Research Office, the Army Research Office through contract no. W911NF-14-1-0679, the Office of Naval Research Young Investigator Program, the NIH through award R01-HD086888 and 1R01HD086888-01, and the National Science Foundation award \#BCS-1441502, \#BCS-1430087 and \#PHY-1554488. The content is solely the responsibility of the authors and does not necessarily represent the official views of any of the funding agencies.

\section*{Author Contributions}

MGM, STS, and DSB developed the experimental paradigm. MGM collected the data. DSB, MGM designed the research. MGM analyzed the data. MGM, STS, and DSB wrote the paper.

\section*{Keywords}

fMRI; functional connectivity; network science; cognitive systems; brain networks; behavioral adaptability; valuation system; visual statistial learning; reinforcement learning

\end{document}